\documentclass[a4paper,fleqn,usenatbib]{mnras}

\usepackage{newtxtext,newtxmath}
\usepackage{subfig}
\usepackage[T1]{fontenc}
\usepackage{ae,aecompl}

\usepackage{graphicx}	
\usepackage{amsmath}	
\usepackage{amssymb}	

\def\ut#1{\mathop{\vtop{\ialign{##\crcr
     $\hfil\displaystyle{#1}\hfil$\crcr\noalign
     {\kern1pt\nointerlineskip}\hbox{$\hfil\sim\hfil$}\crcr
     \noalign{\kern1pt}}}}}

\def\undersymbol#1#2{\mathop{\vtop{\ialign{##\crcr
     $\hfil\displaystyle{#2}\hfil$\crcr\noalign
     {\kern1pt\nointerlineskip}\hbox{$\hfil#1\hfil$}\crcr
     \noalign{\kern1pt}}}}}
\def\arcsec{^{\prime\prime}}
\def\arcmin{^{\prime}}
\def\degr{^\circ}
\def\hour{^{\rm h}}
\def\minute{^{\rm m}}
\def\second{^{\rm s}}

\newcommand{\angstrom}{\mbox{\normalfont\AA}}

\title[DW Cancri in x-rays]{DW Cancri in x-rays}

\author[A.A. Nucita, L. Conversi, \& D. Licchelli]{
A.A. Nucita,$^{1,2}$\thanks{E-mail: nucita@le.infn.it}
L. Conversi,$^{3}$
D. Licchelli, $^{4}$
\\
$^{1}$ Department of Mathematics and Physics {\it ``E. De Giorgi''} , University of Salento, Via per Arnesano, CP-I93, I-73100, Lecce, Italy\\
$^{2}$  INFN, Sezione di Lecce, Via per Arnesano, CP-193, I-73100, Lecce, Italy\\
$^{3}$  Euclid Science Operation Centre, European Space Astronomy Centre, European Space Agency, P.O. Box  78 - 28691, \\
Villanueva de la Canada, (Madrid) - Spain\\
$^{4}$ R.P. Feynman Observatory, Gagliano del Capo, I-73034 Lecce, Italy and \\ 
CBA, Center for Backyard Astrophysics - Gagliano del Capo, I-73034, Lecce, Italy\\
}

\date{Accepted XXX. Received YYY; in original form ZZZ}

\pubyear{2017}

\begin{document}
\label{firstpage}
\pagerange{\pageref{firstpage}--\pageref{lastpage}}
\maketitle

\begin{abstract}
We report on the $XMM$-Newton observation of DW Cnc, a candidate intermediate polar candidate whose historical optical light curve 
shows the existence of periods at $\simeq 38$, $\simeq 86$ and $\simeq 69$ minutes which were interpreted as the white dwarf spin, the orbital and the 
spin-orbit beat periodicities. By studying the $0.3-10$ keV light curves, we confirm the existence of a period at $\simeq 38$ minutes and 
find in the OM light curve a signature for a period at $75\pm 21$ minutes which is consistent with both the orbital and spin-orbit beat. { 
These findings allow us to unveil without any doubt, the nature of DW Cnc as an accreting intermediate polar. The EPIC and RGS source spectra were analyzed and a best fit model, consisting of a multi-temperature plasma, was found. The maximum temperature found when fitting the data is $kT_{max}\simeq 31$ keV which can be interpreted as an upper  limit 
to the temperature of the shock. }
\end{abstract}




\section{Introduction}

{Cataclysmic variables (CVs) are binary systems with a white dwarf (WD) primary star accreting material from a donor (secondary) companion. 
Although the two stars interact principally via the formation of a Roche-lobe (see, e.g. \citealt{Kuulkers2006} for a review), the accretion depends on several 
parameters as the strength of the magnetic field. In fact, depending on its value, CVs can be classified in non-magnetic systems, 
characterized by a weak field ($\ut< 0.1$ MG) and an accretion disk and, possibly, a boundary layer around the primary 
({see, e.g., \citealt{vanteeseling1996,nucita2009,hoard2010,nucita20092,nucita2011,balman2011,nucita2014,mukai2017} to cite a few}); intermediate polars 
(with magnetic field in the range 0.1-10 MG) where the accretion disk is partially disrupted close the central white dwarf 
({see, e.g. \citealt{haberl2002, evans2004a, evans2004b, evans2005, evans2006, evans2007, demartino2004, demartino2005,mukai2015,bernardini2017}}); and 
polars (see, e.g., \citealt{ramsay2004, szkody2004}) consisting of highly magnetized objects ($\ut> 10$ MG) in which the accretion occurs via a mass flow directly 
pushed onto the white dwarf poles.

In intermediate polars and polars the accretion flow (driven by the magnetic field) undergoes a strong shock close to the WD and releases $X$-rays to 
optical emission. Since the magnetic axis is offset from the WD spin one, the observed signal may show a modulation at the spin period and, in some cases, 
at lower time-scales depending whether parts of both poles are visible.   
Sometimes, a modulation in the $X$-rays light curves on the time scale of the orbital period is also observed (\citealt{parker2005}).
These pulsations may be caused by a dependence of the accretion region view on the binary phase. 

Another possibility could be the existence of a second emission component (caused by the interaction of the mass flow with an accretion disk or the white dwarf magnetosphere) 
whose visibility changes with time. In alternative, if a non-axisymmetric disk exists, $X$-rays could suffer of a local absorption when the line of sight intersects the absorption structures. In the latter case, one would expect a decreasing of the modulation depth with increasing $X$-ray energy, thus suggesting photoelectric absorption as the main cause.
	
In this respect, DW Cancri (hereafter DW Cnc) is a variable binary identified as a
CV from its Balmer emission lines (\citealt{stepanian1982,kopylov1988}). \citet{uemura2002} 
reported kilosecond quasiperiodic oscillations (37 and 73 minutes) in the light curve, 
while \citet{rodriguez2004}, by performing radial velocity measurements, showed 
the existence of a period in the range 77-86 minutes. Finally, \citet{patterson2004} reported the results 
of photometric and spectroscopic observations of the target. In particular, strong detection of the periods $\simeq 86.1015(3)$ minutes and
$38.58377(6)$ minutes were derived from radial-velocity measurements and interpreted as, respectively, the orbital period $P_{orb}$ of 
the binary (so that DW Cnc is a candidate CV below the period gap) and the spin period $P_{spin}$ of a magnetic white dwarf. Further analysis on the DW Cnc light curve also showed the existence of a 
strong signal at spin $69.9133(10)$ minutes coinciding with the difference frequency $2\pi/P_{spin} - 2\pi/P_{orb}$. 

\citet{patterson2004} also pointed out that
the detected periods are stable at least over one year and, since the observed light curve resembles the behaviour of 
several members of the DQ Herculis sub-class of CV (i.e. intermediate polar), DW Cnc might be considered as an intermediate polar CV as well.  
As noted by  \citet{patterson2004} (but see also \citealt{mukai2005}), a high energy view of DW Cnc (with a confirmation of the pulse period) was lacking. 
This is of particular interest since the detection of any $X$-ray pulsation at the white dwarf rotational period could be described in the framework of the model
accretion model proposed by \citet{hellier1991} and would be the signature of a channelled accretion in  DW Cnc. 

Hence, we present a $\simeq 9.4$ ks $XMM$-Newton observation of the intermediate polar candidate DW Cnc showing that its spin pulse is clearly detected 
in the X-ray band thus unveiling its nature as a intermediate polar object. We then discuss the spectral and timing analysis conducted 
on data collected by the EPIC and RGS cameras and the OM telescope interpreting the observed properties as possibly due to changes of view of the $X$-ray emitting region.}

\section{{\it XMM}-Newton view of DW Cancri}

\subsection{Data reduction}
\label{datareduction}

DW Cnc (with J2000 coordinates ${\rm RA=07\hour 58\minute 53.10\second}$  and ${\rm DEC = +16\degr 16\arcmin 45.1\arcsec}$) was observed 
by the $XMM$-Newton satellite in 2012 (Observation ID 0673140101) for $\simeq 9.4$ ks during what appeared to be a normal quiescent 
state\footnote{The historical AAVSO (The American Association of Variable Star Observers, https://www.aavso.org/) shows that DW Cnc is stable on a baseline of several years.} 
of the target. The observation started (ended) on $2012/04/02$ at $09:06:03$ ($11:43:04$) UT with the EPIC pn (MOS) camera operating in large (small) window mode. 
The medium filter was selected during the observation. RGS 1 and 2 data were also available.
The Optical Monitor (OM) aboard the satellite also observed the target in fast mode (thus allowing a time resolution of 0.5 seconds) 
and with the UVM2 filter centered at $\simeq 231$ nm. 

The EPIC raw data files (ODFs) were processed using the $XMM$-Science Analysis System (SAS version 17.0.0). The  data
were processed using the latest available calibration constituent
files (CCFs) and the event lists for the three EPIC cameras obtained by running the {\it emchain} and
{\it epchain} task, thus producing calibrated event files.  We then searched for 
segments of the observation affected by soft proton flares and determined a list of good time intervals through which cleaned event files (suitable for the following analysis) were produced. The files were also corrected for the barycenter (via the {\it barycen} SAS tool) so that the photon arrival times are in the barycentric dynamical time instead
of spacecraft time. 

The source (plus background) signals in the soft ($0.3-2$ keV), hard ($2-10$ keV) and full ($0.3-10$ keV) bands were extracted from circular regions centered on the nominal 
DW Cnc coordinates and with radii of $40\arcsec$ thus allowing a collection of $\simeq 88\%$ of the total energy. The background counts 
were extracted (for the same energy bands as above) from surrounding circular regions with radii of $115\arcsec$. For each EPIC camera and energy band, we produced 
synchronized source (plus background) and background light curves with bin size of $10$ seconds. The background light curves were then scaled (mainly accounting for the 
source extraction area) and subtracted from the source light curves by using the {\it epiclccorr} task. Note that the soft, hard and full X-ray light curves were also 
synchronized to each other so that they can be combined thus 
increasing the signal-to-noise ratio. As a result of the above procedure the EPIC soft, hard and full (background corrected) light curves started (ended) at 
$MJD= 56019.40173$ ($MJD= 56019.48471$) and correspond to average count rate of $2.2\pm 0.5$,  $0.7\pm 0.5$, and $2.9\pm 0.7$ counts $s^{-1}$, respectively. 
The soft, hard and full $X$-ray light curves are shown (from top to bottom) in the left panels of Figure \ref{figepic}. 

\begin{figure}
  \centering
\includegraphics[width=0.35\textwidth]{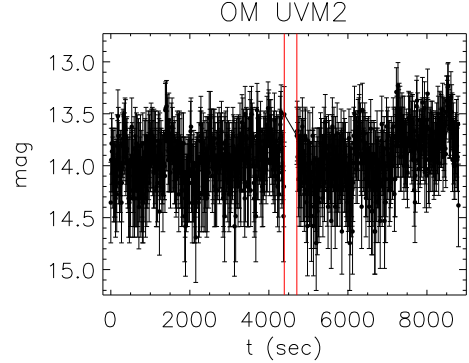}\label{fig3}
\caption{The OM light curve (in the UVM2 filter centered at 231 nm) is shown with a bin size of 10 seconds. The start of the observation corresponds to 
 ${\rm MJD = 56019.38789}$ days with the full light curve lasting for $\simeq 2.44$ hours.}
  \label{figOM}
\end{figure}

The OM UVM2 data where extracted by using the standard {\it omfchain} task with time resolution set to $10$ seconds. This resulted in a count rate light curve  
that was corrected for the solar system barycenter and converted into magnitude assuming a zero point of $\simeq 15.77$. Hence the baseline magnitude 
(see also Figure \ref{figOM} where the start of the observation corresponds to $MJD = 56019.38789$) is $13.93 \pm 0.25$ during the $XMM$-Newton observation. 
{As one can note, the OM observation (lasting for $\simeq 146.6$ minutes) is characterized by a small  gap ($\simeq 318$ seconds) as the  
the maximum allowed integration time for an exposure in fast mode is $\simeq 4.4$ ks. 
}

As far as the spectral data is concerned, the EPIC source and background spectra were extracted in the same regions as above. Furthermore, high resolution spectra from the RGS1 and RGS2  cameras were were obtained (together with the corresponding ancillary files) by running SAS task {\it rgsproc}. Then, the epic source (background corrected) spectra (one for each camera) 
was first re-binned  to  ensure  that  there  were  at  least 25 counts per energy bin and then imported (together with all the relevant quantities 
as the response matrix and ancillary file) within the XSPEC package (version 12.9.0) for the spectral analysis and fitting procedure (see Section \ref{spectral}).

\subsection{Timing analysis}
\label{timing}
\begin{figure*}
  \centering
  \subfloat[]{\includegraphics[width=0.5\textwidth]{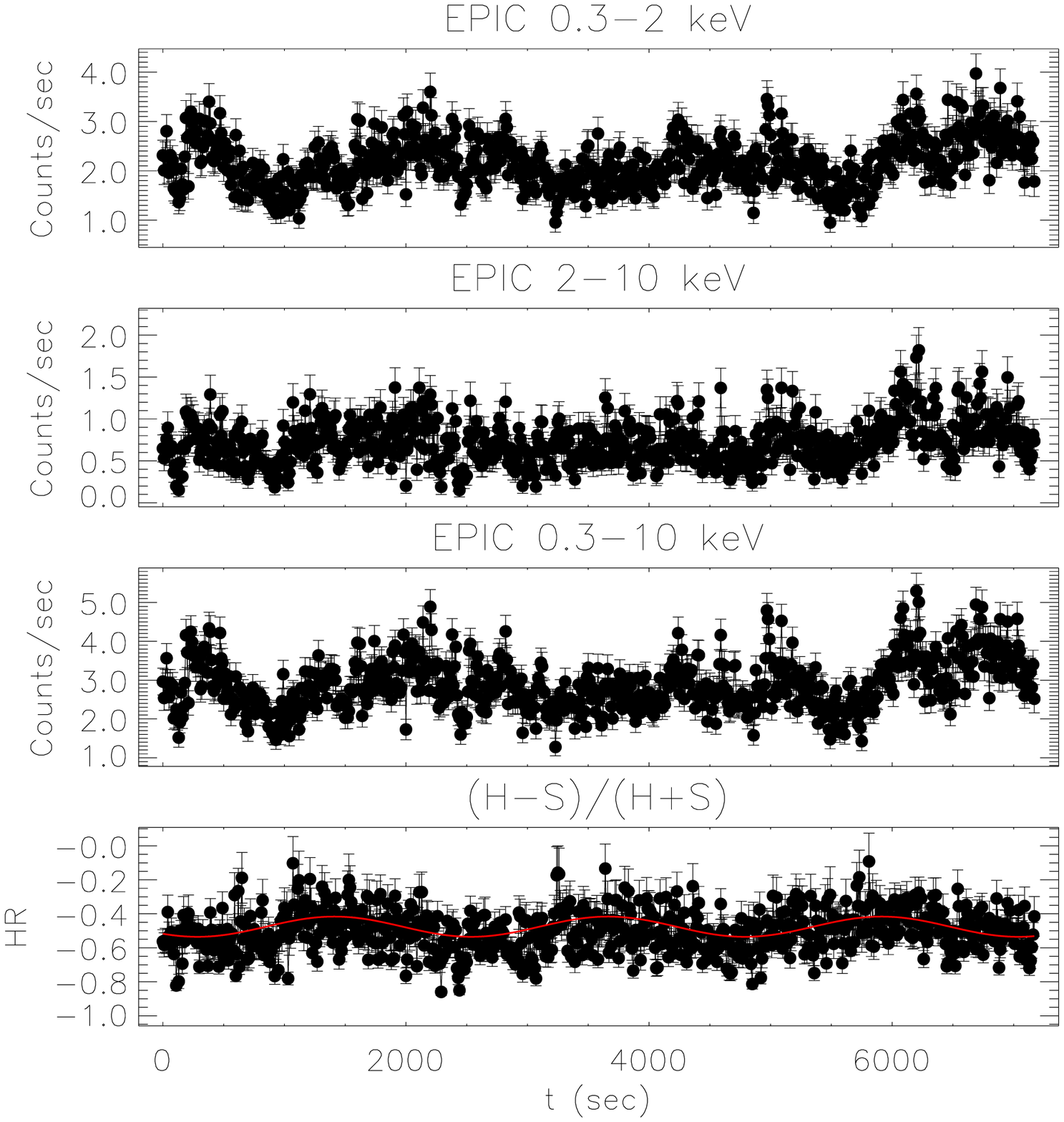}\label{fig1a}}
  \hfill
  \subfloat[]{\includegraphics[width=0.5\textwidth]{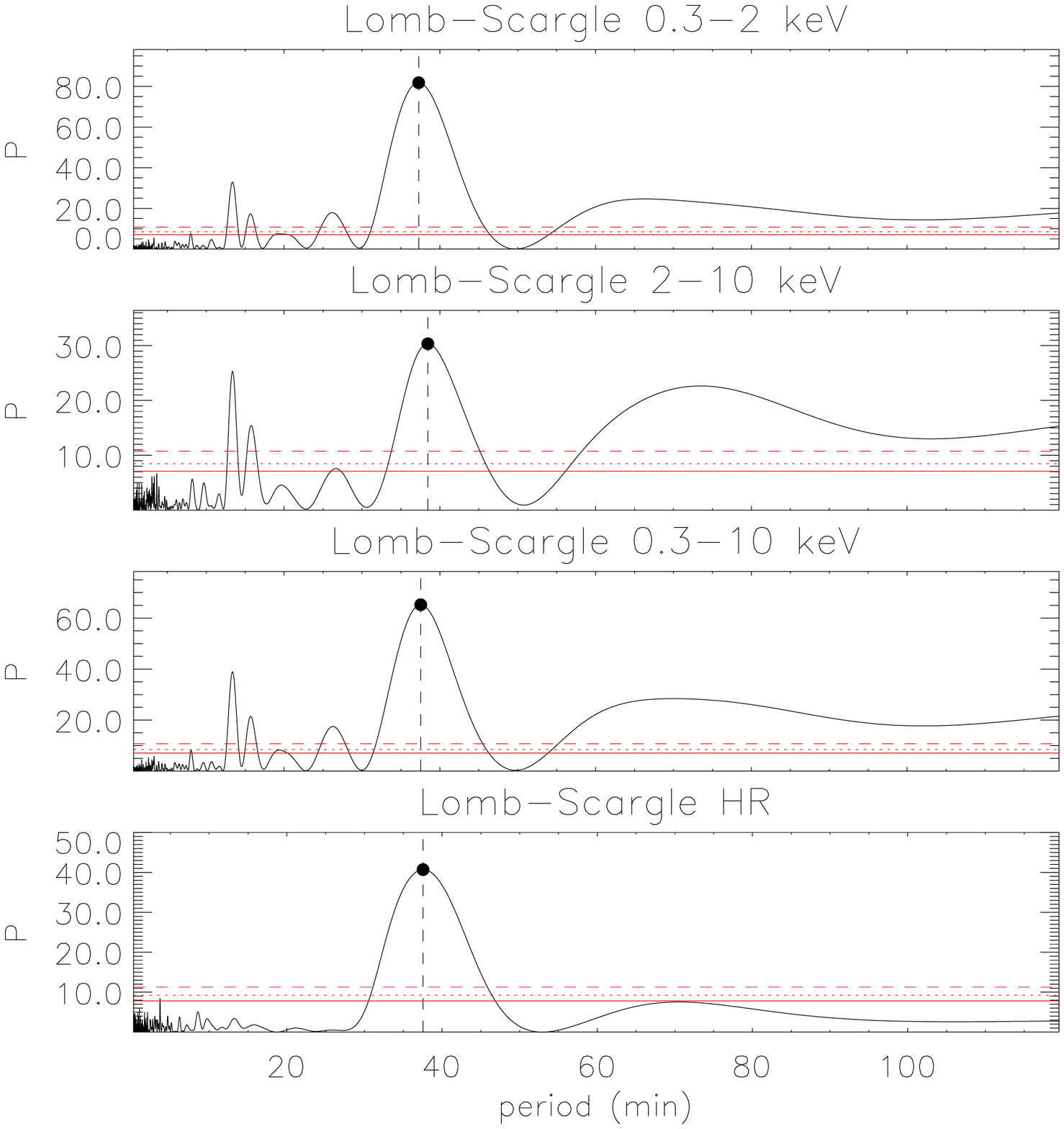}\label{fig1b}}
  \caption{Left panel: the DW Cnc Epic (background subtracted and synchronized) light curves in the 0.3-2 keV, 2-10 keV and 0.3-10 keV bands and
  the hardness ratio together with the best fit sinusoidal function, respectively. 
  Each light curve has a bin size of 10 seconds and the start of the observation corresponds to 
  ${\rm MJD = 56019.40173}$ days. Right panel: the associated Lomb-Scargle periodograms with the identification of the WD spin period.}
  \label{figepic}
  
\end{figure*}

As described in the previous section, the barycentric/background corrected light curves (lasting for $\simeq 119.5$ minutes) 
were extracted in the soft (0.3-2 keV) and hard (2-10 keV) bands with a bin size of $10$ seconds. 
The $X$-ray light curves and the associated hardness ratio were used to perform a blind search for periodicities in the range between twice 
the bin size and half the observational window 
with the well-known Lomb-Scargle technique (\citealt{scargle1982}). This search resulted in the identification of a periodic feature as 
indicated by the dashed vertical lines in the left panels of Figure \ref{figepic}, being the red horizontal lines the false alarm probability thresholds at 68\% (solid), 
90\% (dotted ), and 99\% (dashed) level, respectively. The average position of the feature 
corresponds to the periodicity of $37.7\pm 4.5$ minutes, i.e. consistent with spin period ($P_{spin}$) of the white dwarf already identified by
\citet{uemura2002} and \citet{patterson2004}, thus confirming this signature in the high energy data from DW Cnc. Note that the duration of the $X$-ray 
light curve limits our capability in any clear detection {via the Lomb-Scargle periodogram  technique} of the orbital period (expected to be at $\simeq 86$ minutes) nor 
the periodicity associated to the difference frequency ($\omega_{spin}-\omega_{orb}=2\pi/P_{spin} - 2\pi/P_{orb}$) at $\simeq 69$ minutes. {This is also confirmed when using other analysis tools as the epoch folding technique. We also point to the fact that the timing analysis reveals powers at harmonics of the white dwarf period as, in particular, $P_{spin}/2\simeq 19$ minutes (although at a low confidence level) and $P_{spin}/3\simeq 14$ minutes. More interestingly, we noted in the power spectrum (see Figure \ref{figepic}) the existence of a peak at $\simeq 26$ minutes which resembles the side-band feature ($\omega_{spin}+\omega_{orb}=2\pi/P_{spin} + 2\pi/P_{orb}$) expected from $2\omega_{orb}$ modulation (see, e.g., \citealt{norton1996}). 
}

\begin{figure}
  \centering
  \includegraphics[width=0.5\textwidth]{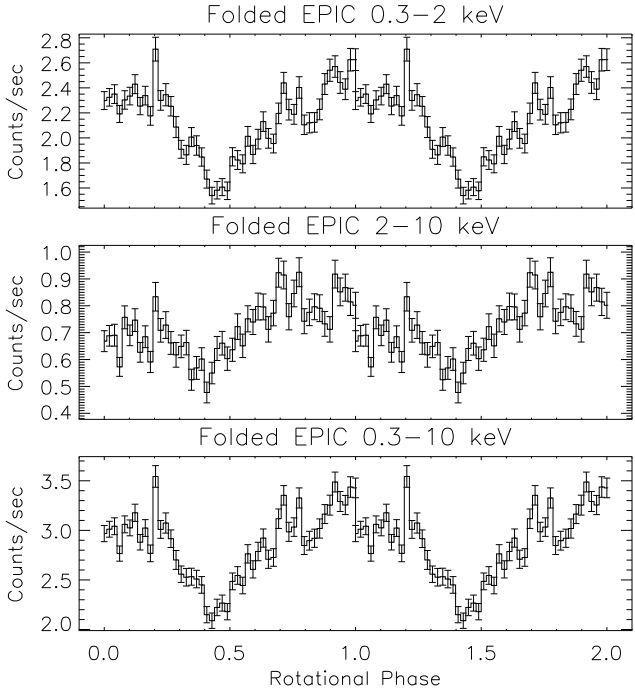}
    \caption{{The EPIC light soft, hard and full curves folded in 50 bins (each corresponding to $\simeq 45.04$ s) at the spin period of the white dwarf.}}
  \label{figepicfolded}
\end{figure}

{We folded the soft, hard and full light curves at the white dwarf spin period (see Figure \ref{figepicfolded}) using 50 bins and note that the data show a 
quasi-sinusoidal behaviour with a broad maximum. In order to evaluate the degree of $X$-ray spin modulation, we define the 
percentage fractional modulation as $100*(M-m)/(M)$ where $M$ and $m$ are the maximum and minimum flux 
in the binned (both soft and hard) light curves. This resulted in fractional modulations resulted to be $43\pm 4\%$  (soft band) and
$48\pm 5\%$ (hard band), respectively. The general smooth and nearly sinusoidal light curves observed in different bands (\ref{figepicfolded}) and the fact that 
the soft and hard fractional modulations are consistent within the quoted errors, possibly imply that the rotational pulses could have their origin in the 
aspect changes (i.e. occultations) of the X-ray emitting region (a polar cap) as the WD rotates
\footnote{Furthermore, since the $X$-ray source never suffers of complete occultations (and no important spectral changes are observed, 
see Section  \ref{spectral}) implies that the oblique rotator is observed at low inclination.}.

Note however that the hardness ratio curve is characterized by a modulation with the white dwarf spin period. This is clear from the bottom panel 
in Figure \ref{figepic} where, as usual, we define the hardness ratio $HR$ as $HR=(H-S)/(H+S)$,
where $S$ and $H$ represents the count rates in the soft and hard bands, respectively. Inspection of this figure shows that the hardness
ratio remains negative (due to an average emission in the soft band larger than the harder one)  
and has a sinusoidal behaviour as shown by the best fit sinusoidal function (obtained fixing the period to that derived 
by the Lomb-Scargle periodogram) superimposed to the data. This behaviour could arise (in part or totally) from the existence of a complex, partial covering absorption 
column as observed in other intermediate polars.  We investigate this issue in Section \ref{spectral} by performing a phase resolved spectral analysis.

{
The OM data in the UVM2 filter covers a total length of $\simeq 146.6$ minutes so that one could in principle detect any 
feature (if present) with periodicity of $\simeq 69-86$ minutes as these signal would be present with at least $\simeq 2$ full cycles. 
The periodicity was searched for by constructing the corresponding Lomb-Scargle periodogram\footnote{We note here that, as discussed in \citet{belanger2016}, 
the presence 
of gaps in the data introduces structures in the periodogram (including the Lomb-Scargle one) and, in particular, 
a sort of reddening effect appears. Hence, 
we filled the OM gap by randomly 
selecting data points from the remaining part of the light curve thus preserving the overall characteristics as bin size, $rms$ and noise of the data. 
For each light curve, the periodicity was searched for by constructing the corresponding Lomb-Scargle periodogram and the results averaged.
}. This procedure resulted in a clear identification in the 
optical data of a period of $75\pm 21$ minutes which, within the quoted errors, is consistent with both the $69$ and $86$ minute 
features previously reported. The large error is due to the peak broadening in the associated periodogram.
}
}

\subsection{Spectral analysis}
\label{spectral}
We first simultaneously fit the background subtracted MOS 1, MOS 2 and pn spectra (grouped on a minimum of 25 counts per energy channel)
with a simple bremsstrahlung model absorbed by a neutral hydrogen foreground column and a constant of
proportionality to account for any possible difference between the detector
responses (within XSPEC the model is  $const*phabs*brems$). We fixed the hydrogen column 
density to the average value found in the direction of the target
($nH =  3.52 \times 10^{20}$ cm$^{-2}$ \citealt{nhtool}).
The resulting fit quality is very poor ($\chi^2=4.05$ for 542  degrees of freedom, d.o.f.) and large residuals appear 
in the energy range $0.7-1$ keV s (possibly due to the -blended- iron L-shell complex around 1 keV) and at the K line 
locations of heavy elements, in particular the features at $6.65$ and $6.69$ keV, plus residuals corresponding to the fluorescence of iron 
at $6.4$ keV. { In this respect, when repeating the analysis by adding three thin Gaussian lines 
(at $6.4$, $6.65$, and $6.69$ keV) and one broad Gaussian centered at $\simeq 0.8$ keV the quality of the fit dramatically improves 
($\chi^2 =1.32$ for 535 d.o.f.)  but still remaining statistically unacceptable. }
\\ These results led us to examine the spectrum with an emission model (as {\it mekal}, \citealt{mekal}) of hot plasma 
in collisional ionization equilibrium able to simulate line emissions from different elements. {A single temperature model 
($const*phabs*mekal$ in XSPEC), with hydrogen column density as above and metal abundances set to the solar ones) resulted in plasma temperature 
of $\simeq 5.3$ keV but very poor from the statistical point of view ($\chi^2=4.83$ for  543 d.o.f.).} 
Since, relaxing the assumption of  the solar abundance did not improve the fit, this is an hint that a multi-temperature plasma is acting. {We found a reasonable 
fit by requiring a three component plasma ($\chi^2=1.05$ for 538 d.o.f.) with  temperatures $kT_1=11.4^{+2.0}_{-2.0}$ keV, 
$kT_2=1.3^{+0.1}_{-0.1}$ keV, and $kT_3=0.60^{+0.10}_{-0.10}$ keV, respectively. The metal abundance  converged towards 
the value $A_Z=0.42\pm 0.10$.   All the errors (here and hereinafter are quoted at the $90\%$ confidence level.

Apart from small residuals around the iron complex lines, we note that
an excess below 0.4 keV is still present. This excess is accounted for when  we relaxed the value of the hydrogen column density so that 
the fit converged towards ($\chi^2=0.97$ for 537 d.o.f.)  a three plasma model with metal abundance of   $A_Z=0.66\pm0.10$ , temperatures $kT_1=11.7^{+2.0}_{-1.0}$ keV, $kT_2=1.3^{+0.1}_{-0.1}$ keV, and $kT_3=0.61^{+0.04}_{-0.04}$ keV and  neutral hydrogen column density $nH =  (1.9\pm 0.4) \times 10^{20}$ cm$^{-2}$. Note that the two previous models converge practically to the same values of temperatures (within the errors) regardless if the hydrogen column density is considered as a free fit parameter or fixed at its average galactic value observed towards the target. 

In the previous Section, we showed that the hardness ratio light curve is characterized by a modulation with the white dwarf spin period. Since most of the IP are characterized by 
strong and very complex absorption  (see, e.g. \citealt{mukai1994}), the observed modulation could be explained by a  partial covering absorption column. 

To test this hypothesis, we fit the data using, as above, three mekal components with equal abundances and absorbed by both simple and partial absorptions, i.e. $const*phabs*pcfabs*(mekal+mekal+mekal)$ in XSPEC.  The best fit resulted in a metal abundance of  $A_Z=0.53^{+0.12}_{-0.09}$, temperatures $kT_1=10.5^{+1.2}_{-1.3}$ keV, $kT_2=1.4^{+0.1}_{-0.1}$ keV, and $kT_3=0.62^{+0.04}_{-0.05}$ keV,  neutral hydrogen column density $nH =  (2.2\pm 0.5) \times 10^{20}$ cm$^{-2}$,  complex equivalent hydrogen column  $nH =  (10.4^{+21.7}_{-9.1}) \times 10^{22}$ cm$^{-2}$ (being this value similar -although with large uncertainties - to column density values found in other IP objects, see e.g. \citealt{evans2006}) and with a partial dimensionless covering factor $f=0.11\pm 0.08$. 

Note that, although all the values of the interesting parameters remain practically unchanged with respect to the previous model, the associated $\chi^2$ statistics ($\chi^2=0.97$ for 535 d.o.f.) and its negligible improvement does not justify the introduction of any complex partial covering  thus favouring galactic absorption.

We further noted that the metal abundance sets to a relatively low value and, in addition, the simple three  temperature plasma model (with all the parameters fixed to their best fit vales except for the normalization constants) does not adapt to the RGS 1 and RGS 2 data failing in reproducing the overall shape and, in particular, the intensities of the transition lines as thw He-like transitions of O{\bf VII} (see Section \ref{rgsspectroscopy}).  Hence, two possibilities are equally probable:  the derived metal abundance is too low or a complex stratification of temperatures is required. 

Observing that the accretion post shock regions are expected to have a gradient in temperature deriving from the cooling of the gas when falling onto the WD surface 
\citep{demartino2005}, we used in XSPEC a multi-temperature scenario based on the {\it cemekl} model (still intrinsically dependent on {\it mekal}) absorbed 
by a neutral hydrogen distribution characterized by a column density. We remind that 
the  {\it cemekl} model is normally used to account for a gradient of temperature in post-shock regions around CVs. In particular, the 
emission measure gradient follows a power law as $d EM/dT =(T/T_{max})^{\alpha-1}/T_{max}$. In this model, the free parameters are the neutral hydrogen 
column density, the maximum value of the plasma temperature $T_{max}$, the power law index $\alpha$, the metal solar abundance $A_Z$ relative to the solar one, the  model normalization and a constant factor introduced for inter-calibration issues among the instruments. The best fit converged towards the values ($\chi^2= 1.04 $ for 540 d.o.f.) 
$nH =  (3.1\pm 0.3) \times 10^{20}$ cm$^{-2}$,  $kT_{max}=31^{+5}_{-4}$ keV, $\alpha=0.65^{+0.06}_{-0.05}$,  and $A_Z=0.9^{+0.1}_{-0.1}$, respectively. 

Note that the low hydrogen column density of the absorber is consistent with the value of the galactic column density in the direction of the source so that, again,
the introduction of any complex partial covering is not required. Differently to what happens in most of the observed IPs (\citealt{mukai1994}), but in common with at least another IP object (HT Cam, \citealt{demartino2005}),  DW Cnc seems to be not characterized by complex local absorption, being this an hint that the X-ray source is not seen though any absorbing material (see also next) possibly in agreement with the fact that the soft and hard fractional modulations are consistent within the quoted errors. 

As a by product, this models reproduces much better the overall structure of the RGS 1 and RGS 2 spectra (see Section \ref{rgsspectroscopy}) when fixing all the parameters to the best fit results and adjusting only the normalizations. 

In order to investigate the white dwarf spin pulse profile and any possible dependence of the source spectral properties on the phase, we defined phases intervals (see Fig. \ref{figepicfolded}) encompassing the regions around maxima (at phases 0-0.25 and 0.7-1) and  minima (at phases 0.33-0.55) and extracted the corresponding spectra 
from the three {\it XMM}-Newton camera. The spectra were then fitted separately with the same model consisting in an absorbed multi-temperature plasma with all the interesting 
parameters free to vary but fixed metal abundance to the value obtained analyzing the phase averaged spectrum. As a result,  we obtained 
$kT_{max}=28^{+6}_{-4}$ keV and $\alpha=0.57^{+0.06}_{-0.05}$, and $kT_{max}=35^{+7}_{-8}$ keV and $\alpha=0.61^{+0.10}_{-0.05}$ for maxima and minima, 
respectively. Clearly,  at different phases, the spectral properties of the source are consistent within the quoted uncertainties. 

The previous finding, and the fact that we do not find evidences for a local partial covering, pushes us in interpreting the observed variability 
due to aspect changes of the underlying $X$-ray emitting region as the WD rotates. In fact, in the scenario in which a polar cap never suffers of complete occultations (i.e. 
the oblique rotator is observed at a low inclination angle), changes in the observed projected area of the $X$-ray source naturally explains the observed modulation.

This model (see Figure \ref{figspectrum_best_fit_cvmekal}) resulted in absorbed and unabsorbed $0.3-10$ keV band fluxes of  $F_{0.3-10 keV}^{abs}= (1.59^{+0.02}_{-0.02})\times 10^{-11}$  ${\rm erg s^{-1} cm^{-2}}$ and 
$F_{0.3-10 keV }^{una}= (1.68^{+0.02}_{-0.02})\times 10^{-11}$ ${\rm erg s^{-1} cm^{-2}}$, respectively, which correspond to luminosities of 
$\simeq 8.2\times 10^{31}$ ${\rm erg s^{-1}}$ and $\simeq 8.7\times 10^{31}$ ${\rm erg s^{-1}}$ when a distance of $\simeq 208$ pc (as reported in the second Gaia data 
release, \citealt{gaiadr2}) is assumed  (see Table \ref{bestfit}).  
}
\begin{table}
	\centering
	\caption{Spectral parameters for the best fitting model ($const*phabs*(cemekl)$ in XSPEC,  see text for details). We also give the $0.3-10$ keV absorbed and unabsorbed fluxes.}
	\label{bestfit}
	\begin{tabular}{l} 
		\hline
                 $nH= (3.1\pm 0.3) \times 10^{20}$~${\rm  cm^{-2}}$  \\ 
                 $\alpha=(0.65^{+0.06}_{-0.5}$\\ 
                 $KT_{max}=31^{+5}_{-4}$ ${\rm keV}$s   \\ 
                 $A_Z=0.9^{+0.1}_{-0.1}$ \\ 
		 $F_{0.3-10 keV}^{abs}= (1.59^{+0.02}_{-0.02})\times 10^{-11}$  ${\rm erg ~cm^{-2} ~s^{-1} }$\\
		 $F_{0.3-10 keV }^{una}= (1.68^{+0.02}_{-0.02})\times 10^{-11}$ ${\rm erg ~cm^{-2} ~s^{-1} }$\\
 		 \hline
	\end{tabular}
\end{table}
\begin{figure*}
  \centering
  \includegraphics[width=0.5\textwidth, angle=-90]{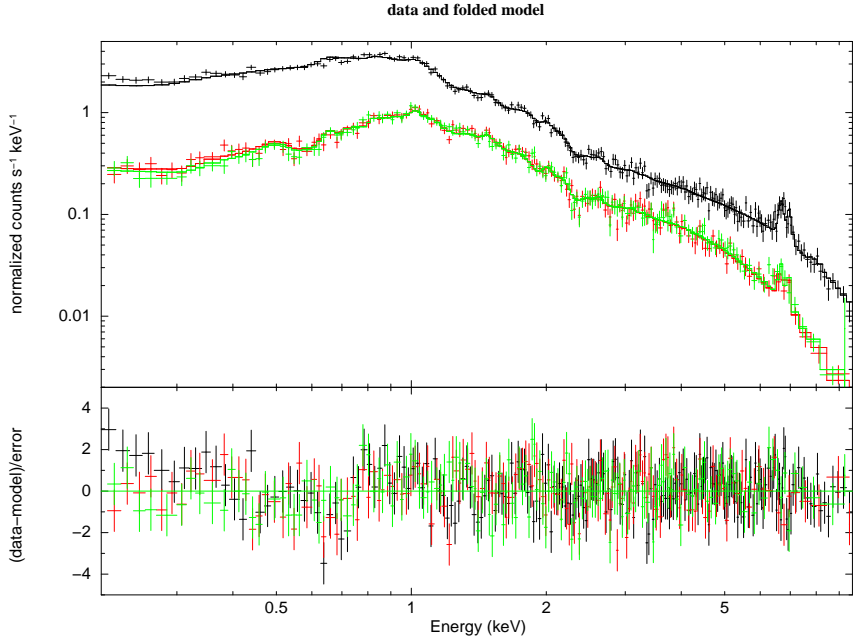}
  \caption{The best fit to the MOS 1, MOS 2, and pn data with the XSPEC model $const*phabs(cemekl)$ (see text for details).}
  \label{figspectrum_best_fit_cvmekal}
\end{figure*}

\subsection{RGS data}
\label{rgsspectroscopy}

\begin{figure*}
  \centering
  \includegraphics[width=0.5\textwidth, angle=-90]{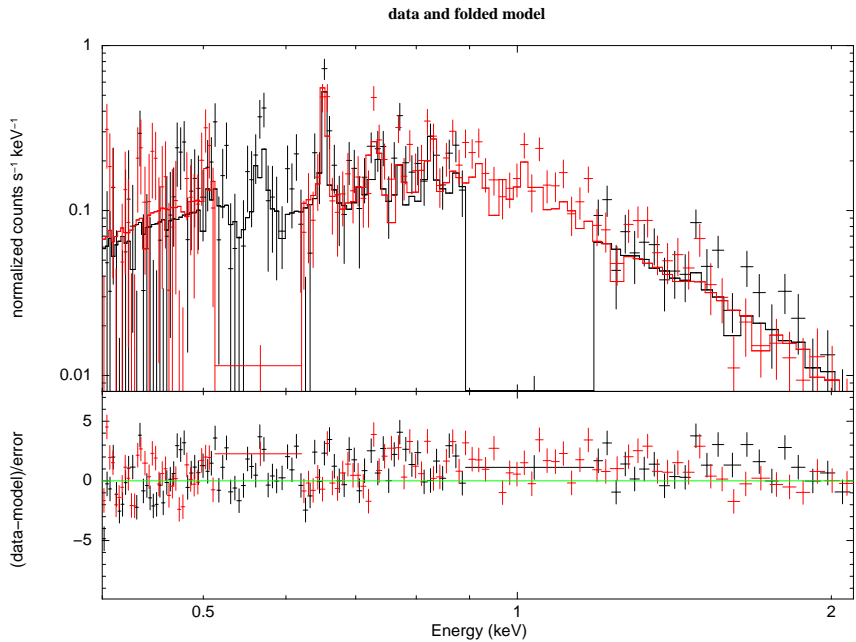}
  \caption{The best fit model superimposed on the RGS 1 (red) and RGS 2 (black) data. Note the presence a strong O{\bf VIII} Ly-$\alpha$ line at $18.96$ $\angstrom$ 
  and the O{\bf VII} He-like triplet.}
  \label{figspectrumfitbbrgs}
\end{figure*}

Note that the same multi-temperature model used for the EPIC spectral fit provides an adequate description of the RGS spectra  when fixing all the parameters of the model to those
derived above but adjusting consistently the model normalizations. In particular, inspecting Figure \ref{figspectrumfitbbrgs} (where the best fit is superimposed to the RGS1 and RGS2 data re-binned in order to have at least a significance of 10 $\sigma$ per bin), one can note the existence of the O{\bf VIII} Ly-$\alpha$ emission line at $\simeq 0.65$ keV, and
emission lines in the energy range 0.56 keV-0.57 keV (21.6 $\angstrom$-22.1 $\angstrom$) possibly associated to the 
He-like transitions of O{\bf VII}, i.e. the resonance line $r$ (corresponding to the transitions between the $n=2$ shell and the $n=1$ ground state), and the 
inter-combination ($i$) and the forbidden ($f$) lines. In this respect, 
as shown by \citet{porquet2000} the relative emission strength of the $r$, $i$ and $f$ lines is a good indicator of the physical conditions of density and temperature of the gas. 
\\In order to have a measure of the line fluxes, we followed the phenomenological spectral analysis method described in \citet{nucita2010} and references therein.
In particular, the unbinned RGS spectra are divided in intervals of 100 channels wide and Gaussians are fitted to all the identified emission lines. 
For each (single) emission line the centroid energy is free to vary and, when triplets are identified, the relative distance between the central energies was frozen to the value
predicted by atomic physics. The local continuum was always modelled as a power law with a fixed photon index
$\Gamma =1$ and normalization free to vary. Since we are dealing with the unbinned spectra, we estimated the goodness of the fit by using the 
C-statistic (see, \citealt{cash1979}). We further consider the line (or triplet) detected at $68\%$ confidence level when, repeating the fit with the continuum only, we obtained a change in the C-statistic value ($\Delta C$) by at least 2.3. The result of this analysis (see also Table \ref{tabvleOVII}) showed that the O{\bf VII} resonance line is much more weaker than the inter-combination line so that ionization processes may occur in the medium (\citealt{porquet2000}). This is also clear when using the standard
line ratios $R=f/i$, $L=r/i$, and $G=(f+i)/r$ as standard diagnostic. In particular, as clear from table \ref{tabvleOVII}, the large (small) values of the $G$ ($R$) ratio imply 
photonionization and, since $R$ decreases with increasing values of the electron density $n_e$, a lower limit on the electron density of $\simeq 10^{11}$ cm$^{-3}$ (\citealt{porquet2000}).
\begin{table*}
	\centering
	\caption{Line parameters as measured from the RGS 1 and RGS 2 spectra of DW Cnc. The expected centroid wavelength in the rest-frame are extracted from 
	the CHIANTI database (\citealt{chianti}).} 
	\label{tabvleOVII}
	\begin{tabular}{llllllll} 
        \hline
         {\rm Line ID} & $\lambda_{exp}($\angstrom$)$ & $\lambda_{obs}($\angstrom$)$ & Flux($\times 10^{14}$ erg s$^{-1}$ cm$^{-2}$)  & $\Delta C$ & R & L & G \\
        \hline
         O{\bf VIII} Ly-$\alpha$ & $18.969$ & $18.966^{+0.009}_{-0.012}$  & $13.6^{+0.1}_{-0.1}$  &  $16$   &  -- & -- & -- \\
         O{\bf VII} (r)          & $21.600$  & $21.632^{+0.019}_{-0.008}$ & $5.5^{+3.2}_{-2.9}$  &  $14$ &  $\ut<1$ & $\ut < 1.5$ & $\ut<5$ \\
         O{\bf VII} (i)          & $21.790$  & $21.835^{+0.019}_{-0.008}$ & $8.1^{+4.7}_{-4.6}$  &  --  &  -- & -- & -- \\
         O{\bf VII} (f)          & $22.101$  & $22.135^{+0.019}_{-0.008}$ & $3.0^{+3.4}_{-2.6}$  &  --   &  -- & -- & -- \\
        \hline
	\end{tabular}
\end{table*}
Note also that the best fit line centroids seem to be shifted compared to the laboratory values by  $-40^{+150}_{-190}$ km s$^{-1}$ for the O{\bf VIII} Ly-$\alpha$ line and 
$440^{+260}_{-100}$ km s$^{-1}$ for the O{\bf VII}  triplet. However, due to the large uncertainties, further higher S/N observations are required in 
order to accurately determine the redshift and, possibly, to analyze any line modulation with the spin/orbital period of DW Cnc.

\section{Discussion and results}
\label{discussion}
{
DW Cancri is a variable binary classified as a CV from its Balmer emission lines (\citealt{stepanian1982,kopylov1988}). The source was intensively studied 
in the optical band and periods of $\simeq 38$, $\simeq 86$ and $\simeq 69$ minutes (see, e.g., 
\citealt{uemura2002}, \citealt{rodriguez2004}, \citealt{patterson2004}) were identified and explained as the WD spin period, the orbital period and spin-orbit 
beat, respectively. 

Nevertheless, as suggested by \citet{patterson2004},  a confirmation of the DW Cnc spin period in the $X$-rays was necessary.  
Hence, analyzing the 0.3-10 keV data acquired by the $XMM$-Newton telescope, we have shown the existence of a $37.7\pm 4.5$ minute periodicity consistent with the 
spin period of the white dwarf thus allowing us to classify DW Cnc a member of the intermediate polar class. The timing analysis reveals powers at harmonics 
of the white dwarf period, i.e $P_{spin}/2\simeq 19$ minutes and $P_{spin}/3\simeq 14$ minutes, being the second feature more evident.
Unfortunately, the duration of the $X$-ray light curve limited our capability to have any clear detection of the orbital and spin-orbit beat periodicity corresponding to the frequency $\omega_{spin}-\omega_{Orb}=2\pi/P_{spin} -2\pi/P_{orb}$. However, a sideband at the beat period $\simeq 26$ minutes possibly exists as originating from a $2\omega_{orb}$ modulation which could produce a feature at frequency $\omega_{spin}+\omega_{Orb}=2\pi/P_{spin} + 2\pi/P_{orb}$. 

Note finally that  the OM data in the UVM2 filter covers a total length of $\simeq 146.6$ minutes thus allowing us to identify  a period of $75\pm 21$ minutes which, within the quoted uncertainty, is consistent with both the $69$ and $86$ minute features reported in literature. 

We also performed a spectral analysis of the phase average spectrum finding that the EPIC data can be described by a multi-temperature plasma model simply 
absorbed by the galactic neutral hydrogen column density. In particular, the best fit converged towards the values ($\chi^2= 1.04 $ for 540 d.o.f.) 
$nH =  (3.1\pm 0.3) \times 10^{20}$ cm$^{-2}$,  $kT_{max}=31^{+5}_{-4}$ keV, $\alpha=0.65^{+0.06}_{-0.05}$,  and $A_Z=0.9^{+0.1}_{-0.1}$, respectively which 
corresponds the unabsorbed  $0.3-10$  keV band flux of  $F_{0.3-10 keV }^{una}= (1.68^{+0.02}_{-0.02})\times 10^{-11}$ ${\rm erg s^{-1} cm^{-2}}$. When assuming 
a distance of  $\simeq 208$ pc, the measured luminosity turns out to be $\simeq 8.7\times 10^{31}$ ${\rm erg s^{-1}}$.

The best fit model provides an adequate description of the RGS spectra  ($<0.3$ keV) as well. In addition, in  the low energy data, we noted the existence of 
a few emission lines (corresponding to the  O{\bf VIII} Ly-$\alpha$ transition and the O{\bf VII} He-like triplet) which give us an hint  that a photonionization mechanism is acting. In particular, the estimated large (small) values of the $G$ ($R$) ratios (associated to the  the O{\bf VII} He-like triplet) suggests
photonionization and, since $R$ decreases with increasing values of the electron density $n_e$, a lower limit on the electron density of $\simeq 10^{11}$ cm$^{-3}$ (\citealt{porquet2000}).

We note here that, since $XMM$-Newton data extends only up to $\simeq 10$ keV, a plasma temperature as large as $kT\simeq 31$ keV (corresponding to the harder part of the {\it mekal} components embedded in the {\it cvmekal} model used here) could be un-reliable (but not different from other similar findings as in the case of PQ Gem \citep{evans2006})
and possibly arising from the usage of a multi-temperature model. Of course, further spectral observations extended to larger energies would allow to solve this issue.

We also remind  that the spectral properties extracted in the phase intervals associated to the maxima and the minima of the source activity remain consistent (within the errors) 
with the values derived for the phase averaged spectrum. This fact, together with the absence of any evidence of a local partial covering, push us in interpreting the observed variability as due to aspect changes of the X-ray emitting region as the WD rotates. Also in this case, further and longer $X$-ray  and/or optical observations are 
required in order to check the existence of periods other than that associated to the white dwarf spin in the DW Cnc data. High energy band X-ray data (see e.g. \citealt{landi}) would also allow to firmly establish the spectral properties of DW Cnc.}

\section*{Acknowledgements}
This  paper  is  based on observations from
XMM-Newton, an ESA science mission with instruments
and  contributions  directly  funded  by ESA Member  States  and  NASA. 
We thank the anonymous Referee for the suggestions that greatly improved the paper. 
We thank for partial support the INFN projects TAsP and EUCLID. We warmly acknowledge Berlinda Maiolo, 
Sara Nucita, and Matteo Nucita for reading the manuscript. We warmly acknowledge ESAC (ESA) for
the facilities provided. 






\bsp	
\label{lastpage}

\begin{thebibliography}{99}


\bibitem[\protect \citeauthoryear{Brown et al.}{2018}]{gaiadr2}
A. G. A. Brown, A. Vallenari, T. Prusti, J. H. J. de Bruijne, et al. A\&A, 2018

{\bibitem[\protect \citeauthoryear{Balman}{2011}]{balman2011}
Balman, S., 2011, ApJ, 741, 84}


{\bibitem[\protect \citeauthoryear{Belanger}{2016}]{belanger2016}
Belanger, G., 2016, ApJ, 822, 14}

{\bibitem[\protect \citeauthoryear{Bernardini et al.}{2017}]{bernardini2017}
Bernardini, F., et al., 2017, MNRAS, 470, 4815B}


\bibitem[\protect \citeauthoryear{Cash}{1979}]{cash1979}
Cash, W. 1979, ApJ, 228, 939

\bibitem[\protect \citeauthoryear{Dere}{2001}]{chianti}
Dere, K. P., 2001, ApJSS, 134, 331

{\bibitem[\protect \citeauthoryear{de Martino et al.}{2004}]{demartino2004}
de Martino, D., Matt, G., Belloni, T., Haberl, F., \& Mukai, K.,  2004, A\&A, 415, 1009D}

{\bibitem[\protect \citeauthoryear{de Martino et al.}{2005}]{demartino2005}
de Martino, D., et al., 2005, A\&A, 437, 935D}


{
\bibitem[\protect \citeauthoryear{Evans \& Hellier}{2004 a}]{evans2004a}
Evans, P.A., \& Hellier, C., 2004, MNRAS, 353, 447 
}

{
\bibitem[\protect \citeauthoryear{Evans et al.}{2004 b}]{evans2004b}
Evans, P.A., Hellier, C., Ramsay, G., \& Cropper, M., 2004, MNRAS, 349, 715 
}


{
\bibitem[\protect \citeauthoryear{Evans \& Hellier}{2005}]{evans2005}
Evans, P.A., \& Hellier, C., 2005, MNRAS, 359, 1531 
}

{
\bibitem[\protect \citeauthoryear{Evans et al.}{2006}]{evans2006}
Evans, P.A., Hellier, C., \& Cropper, M., 2006, MNRAS, 369, 1229 
}

{
\bibitem[\protect \citeauthoryear{Evans \& Hellierr}{2007}]{evans2007}
Evans, P.A., \& Hellier, C., 2007, ApJ, 663, 1277 
}

{\bibitem[\protect \citeauthoryear{Haberl}{2002}]{haberl2002}
Haberl., F., 2002, in {\it The Physics of Cataclysmic Variables and Related Objects},
ASP Conference Proceedings, Vol. 261. Eds.B. T. Gänsicke, K. Beuermann, \& K. Reinsch}

{\bibitem[\protect \citeauthoryear{Hellier et al.}{1991}]{hellier1991}
Hellier, C., Cropper, M., \& Mason, K.O., 1991, MNRAS, 248, 233
}


{\bibitem[\protect \citeauthoryear{Hoard et al.}{2010}]{hoard2010}
Hoard, D.W., et al., 2010, AJ, 140, 1313
}




\bibitem[\protect \citeauthoryear{Kalberla et al.}{2005}]{nhtool}
Kalberla, P. M. W.,  et al. 2005, A\& A, 440, 775

\bibitem[\protect \citeauthoryear{Kopylov et al.}{1988}]{kopylov1988}
Kopylov, I. M., et al. 1988, Astrofizika, 28, 287

\bibitem[\protect \citeauthoryear{Kuulkers et al.}{2006}]{Kuulkers2006}
Kuulkers, E., Norton, A., Schwope, A., \& Warner B., 2006, in Compact Stellar
X-ray  Sources,  ed.  W.  H.  G.,  Lewin,  \&  M.,  van  der  Klis,  Cambridge
Astrophys. Ser., 39, 421


\bibitem[\protect \citeauthoryear{Landi et al.}{2009}]{landi}
Landi, R., et al., 2009, MNRAS, 392, 630


\bibitem[\protect \citeauthoryear{Mewe et al.}{1985}]{mekal}
Mewe, R., Gronenschild, E. H. B. M., \& van den Oord, G. H. J. , 1985, A\&AS , 62, 197


{
\bibitem[\protect \citeauthoryear{Mukai et al.}{1994}]{mukai1994}
Mukai, K., , Ishida, M., \& Osborne, J.P., 1994, PASJ, 46, L87
}


	
\bibitem[\protect \citeauthoryear{Mukai}{2005}]{mukai2005}
Mukai, K., 2005, in {\it The Astrophysics of Cataclysmic Variables and Related Objects},
Proceedings of ASP Conference Vol. 330. Edited by J.-M. Hameury and J.-P. Lasota. San Francisco: Astronomical Society of the Pacific, p.147

{
\bibitem[\protect \citeauthoryear{Mukai et al.}{2015}]{mukai2015}
Mukai, K., Rana, V., Bernardini, F., \& de Martino, D., 2015, ApJ, 807L, 30M

\bibitem[\protect \citeauthoryear{Mukai et al.}{2017}]{mukai2017}
Mukai, K., 2017, PASP, 2017, 129

}




{
\bibitem[\protect \citeauthoryear{Norton et al.}{1996}]{norton1996}
Norton, A.J., Beardmore, A.P., \& Taylor, P., 1996, 280, 3, 937 
}


\bibitem[\protect \citeauthoryear{Nucita et al.}{2009 a}]{nucita2009}
Nucita, A.A., et al., 2009, New Ast,14, 302N

\bibitem[\protect \citeauthoryear{Nucita et al.}{2009 b}]{nucita20092}
Nucita, A.A., et al., 2009, A\&A, 504, 973N

\bibitem[\protect \citeauthoryear{Nucita et al.}{2011}]{nucita2011}
Nucita, A.A., et al., 2011, A\&A, 536, 75N

\bibitem[\protect \citeauthoryear{Nucita et al.}{2014}]{nucita2014}
Nucita, A.A., et al., 2014, A\&A, 566, A121 

\bibitem[\protect \citeauthoryear{Nucita et al.}{2010}]{nucita2010}
Nucita, A.A., et al., 2010, A\&A 515, A47

\bibitem[\protect \citeauthoryear{Patterson et al.}{2004}]{patterson2004}
Patterson, J., et al., 2004, PASP, 116, 516

{
\bibitem[\protect \citeauthoryear{Parker et al.}{2005}]{parker2005}
Parker, T.L., Norton, A.J., \& Mukai, K., A\&A, 439, 213
}

\bibitem[\protect \citeauthoryear{Porquet \& Dubau}{2000}]{porquet2000}
Porquet, D., \& Dubau, J. 2000, A\&AS, 143, 495

{
\bibitem[\protect \citeauthoryear{Ramsay et al.}{2004}]{ramsay2004}
Ramsay, G., et al., 2004, MNRAS, 350, 1373
}

\bibitem[\protect \citeauthoryear{Rodriguez-Gil et al.}{2004}]{rodriguez2004}
Rodriguez-Gil, P., Gaensicke, B. T., Araujo-Betancor, S., \& Casares,
J. 2004, MNRAS, 349, 367

\bibitem[\protect \citeauthoryear{Scargle}{1982}]{scargle1982}
Scargle, J. D. 1982, ApJ, 263, 835

{
\bibitem[\protect \citeauthoryear{Szkody et al.}{2004}]{szkody2004}
Szkody, P., et al., 2004, AJ, 128, 2443S
}

\bibitem[\protect \citeauthoryear{Stepanian}{1982}]{stepanian1982}
Stepanian, J. A., 1982, Perem. Zvesdy, 21, 691


\bibitem[\protect \citeauthoryear{Uemura}{2002}]{uemura2002}
Uemura, Y., et al. 2002, PASJ, 54, 299

\bibitem[\protect \citeauthoryear{van Teeseling et al.}{1996}]{vanteeseling1996}
van Teeseling, A, Beuermann, K., \& Verbunt., F., 1996, A\&A, 315, 467





\end{thebibliography}
\end{document}